\documentclass[journal=./achemso/jpc,manuscript=article,layout=onecolumn]{achemso}
\usepackage{comment}
\usepackage{amsmath}
\usepackage{amsfonts,amstext,amsgen,amsbsy,amsopn,amssymb}

\usepackage{graphicx}
\usepackage{subcaption}
\usepackage{dcolumn}
\usepackage{bm}
\usepackage{physics}
\usepackage{makecell}
\usepackage{multirow}
\usepackage{tabstackengine}
\usepackage{qcircuit}
\stackMath
\usepackage[dvipsnames, table]{xcolor}
	\definecolor{diagramColor}{RGB}{0,0,120}
\usepackage{tikz}
	\usetikzlibrary{trees}
	\tikzset{
 		treenode/.style = {shape=rectangle, draw, diagramColor, ultra thick, align=center, 
                     top color=white, bottom color=white},
		root/.style     = {treenode, font=\normalsize},
  		env/.style      = {treenode, font=\normalsize},
		envtwo/.style      = {treenode, font=\normalsize},
		mode/.style = {edge from parent path={(\tikzparentnode.east) -- (\tikzchildnode.west)}}
	}
\usepackage{enumitem}
\usepackage{mathtools}

\usepackage[colorlinks=true,linkcolor=blue,citecolor=blue,urlcolor=black]{hyperref}
\usepackage{array}
\newcolumntype{?}{!{\vrule width 1.5pt}}
\usepackage{arydshln}

\newcommand{\halfcheckmark}[0]{\checkmark\raisebox{0.23em}{\kern-0.68em\large$\times$}}

\SectionNumbersOn

\title{Mapping Molecular Hamiltonians into Hamiltonians of Modular cQED Processors}

\author{Ningyi Lyu}
\affiliation{Department of Chemistry, Yale University, New Haven, CT 06520, U.S.A.}
\author{Alessandro Miano}
\affiliation{Department of Applied Physics and Physics, Yale University, New Haven, CT 06520, U.S.A.} 
\alsoaffiliation{Yale Quantum Institute, Yale University, New Haven, CT 06511, U.S.A.}
\author{Ioannis Tsioutsios}
\affiliation{Department of Applied Physics, Yale University, New Haven, CT 06520, U.S.A.} 
\alsoaffiliation{Yale Quantum Institute, Yale University, New Haven, CT 06511, U.S.A.}
\author{Rodrigo Corti\~{n}as}
\affiliation{Department of Applied Physics, Yale University, New Haven, CT 06520, U.S.A.} 
\alsoaffiliation{Yale Quantum Institute, Yale University, New Haven, CT 06511, U.S.A.}
\author{Kenneth Jung}
\affiliation{Department of Chemistry, Yale University, New Haven, CT 06520, U.S.A.}
\altaffiliation{Current Address: Department of Chemistry, Stanford University, Stanford, CA 94305, U.S.A.}
\author{Yuchen Wang}
\affiliation{Department of Chemistry, Department of Physics and Purdue Quantum Science and Engineering Institute, Purdue University, West Lafayette, Indiana 47907, USA}
\author{Zixuan Hu}
\affiliation{Department of Chemistry, Department of Physics and Purdue Quantum Science and Engineering Institute, Purdue University, West Lafayette, Indiana 47907, USA}
\author{Eitan Geva}
\affiliation{Department of Chemistry, University of Michigan, Ann Arbor, MI 48109, USA}
\author{Sabre Kais}
\affiliation{Department of Chemistry, Department of Physics and Purdue Quantum Science and Engineering Institute, Purdue University, West Lafayette, Indiana 47907, USA}
\author{Victor S. Batista}
\affiliation{Department of Chemistry, Yale University, New Haven, CT 06520, U.S.A.}
\alsoaffiliation{Yale Quantum Institute, Yale University, New Haven, CT 06511, U.S.A.} 
\email{victor.batista@yale.edu}

\date{\today}

\begin{document}


\abstract{We introduce a general method based on the operators of the Dyson-Masleev transformation to map the Hamiltonian of an arbitrary model system into the Hamiltonian of a circuit Quantum Electrodynamics (cQED) processor.  Furthermore, we introduce a modular approach to program a cQED processor with components corresponding to the mapping Hamiltonian. The method is illustrated as applied to quantum dynamics simulations of the Fenna-Matthews-Olson (FMO) complex and the spin-boson model of charge transfer. Beyond applications to molecular Hamiltonians, the mapping provides a general approach to implement any unitary operator in terms of a sequence of unitary transformations corresponding to powers of creation and annihilation operators of a single bosonic mode in a cQED processor.}

\section{Introduction}
The development of quantum computing simulations for modeling chemical systems is a subject of immense interest. Recent studies have already  explored the potential of quantum computing as applied to electronic structure calculations,\cite{Cao2019,Lee2019,Armaos2020,Xia2021,peruzzo2014variational,o2016scalable,xia2017electronic,xia2018quantum}
quantum dynamics simulations\cite{wang2022simulation,wiebe2011simulating,ollitrault2021molecular,yao2021adaptive,tagliacozzo2022optimal,
wang2011quantum,wei2016duality,kliesch2011dissipative,sweke2015universal,schlimgen2021quantum,zhang2022quantum}
as well as simulations of molecular spectroscopy.\cite{Lee2021,Lee2022a,Lee2022b,Parrish2019} 
Currently quantum computing facilities are often called noisy intermediate-scale quantum (NISQ) computers,\cite{Preskill2018} due to their intrinsic limitations, including architectures based on superconducting circuits,\cite{OMalley2016} trapped ions,\cite{Shen2018,Hempel2018} and nuclear magnetic resonance \cite{Du2010,Li2011}.
To achieve moderate accuracy and reliability in spite of noise and decoherence, simulations of chemical systems have relied on hybrid quantum-classical algorithms, including the variational quantum eigensolver (VQE) method\cite{Colless2018,Lee2021,Kandala2017} and quantum machine learning methods\cite{khalid2022finite,sajjan2022quantum} where only part of the computation is performed on the quantum computer, sometimes applied with the aid of error mitigation techniques,\cite{tazhigulov2022simulating} while the rest of the calculation is run on a conventional computer.

New hardware settings that can fundamentally mitigate the aforementioned errors of quantum computing architectures are necessary to enable fault-tolerant quantum computations of chemical systems. A promising paradigm-shifting technology involves the development of bosonic circuit Quantum Electrodynamics (cQED) processors where information is stored as microwave photons in the unbounded Hilbert space of superconducting oscillator modes. The non-linearity necessary for control and readout procedures is provided by quantum circuits based on ancillary Josephson junctions.\cite{Joshi2021,Blais2021} Bosonic cQED devices offer favorable platforms for quantum error correction codes as a result of the well understood dominant source of errors in oscillator modes, namely, the single-photon loss.~\cite{sivak2022real} Moreover, encoding information in multiple levels of an oscillator can be more efficient when compared to conventional cQED architectures where the storage of information utilizes only the first two-levels of a transmon. 



cQED bosonic devices have already been shown to offer unparalleled capabilities for simulations of vibronic spectra of small molecules such as water, ozone, nitrogen dioxide and sulfur dioxide, when mapping the calculation of Franck-Condon factors into a Gaussian boson sampling problem.\cite{Wang2020} The corresponding calculations on a conventional quantum computer would require 8 qubits and $\mathcal{O}(10^3)$ gates, exceeding the capabilities of current technologies. Therefore, it is natural to anticipate that cQED bosonic devices could be applied to solve other classes of interesting problems in chemistry and offer advantages beyond the capabilities of conventional quantum computers. However, a general approach to design a quantum circuit to simulate an arbitrary molecular system has yet to be established. Here, we address the fundamental question regarding how to map the Hamiltonian of a molecular system into the corresponding Hamiltonian of a programmable cQED bosonic simulator. We introduce the single-bosonic-mode (SBM) mapping, allowing us to represent any square matrix as a polynomial of powers of creation and annihilation operators of a bosonic mode. The mapping thus provides a general protocol for transforming any Hamiltonian into the Hamiltonian of a cQED device, since the Hamiltonian of a cQED device can be written as a polynomial of powers of creation and annihilation operators of a single bosonic mode\cite{Wang2020}. Additionally, we introduce a modular approach to program a cQED processor according to the SBM mapping Hamiltonian. In particular, we identify circuits with Superconducting Nonlinear Asymmetric Inductive eLements (SNAILs)\cite{Zorin2016,Frattini2017,Zorin2021} that could be coupled by beam-splitters, or by nearly-quartic elements\cite{ye2021engineering} for programming one-qubit gates and the two-qubit controlled-Z gate that enable universal computing. 




We illustrate the SBM mapping in conjuction with SNAIL gates as applied to model simulations of quantum dynamics in the photosynthetic Fenna-Matthews-Olson (FMO) complex, a system that mediates the excitation energy transfer from light-harvesting chlorosomes to the bacterial reaction center. Additionaly, we illustrate the SBM mapping as applied to simulations of charge or energy transfer processes with dissipation according to the spin-boson model. Beyond applications to molecular Hamiltonians, the SBM mapping provides a general approach for implementing any unitary operator in terms of a sequence of unitary transformations corresponding to powers of creation and annihilation operators of single-bosonic modes in a cQED processor.




The paper is organized as follows. Section \ref{sec:sbm} introduces the SBM mapping method. Section \ref{sec:SNAIL} provides the implementation of one-qubit gates with capacitively shunted SNAILs, and the two-qubit controlled-Z gate with nearly-quartic elements. Section \ref{sec:models} demonstrates the SBM mapping with SNAIL circuit implementation as applied to quantum dynamics simulations of a series of models typically employed to simulate charge and energy transfer processes.  Conclusions are outlined in Section \ref{sec:conclusion}.

\section{Single-bosonic mode mapping}\label{sec:sbm}
The SBM mapping transforms an arbitrary Hermitian operator,
\begin{equation}\label{Hmn}
\hat{H}=\sum_{\alpha=0}^{k-1}\sum_{\alpha'=0}^{k-1}H_{\alpha\alpha'}|{\alpha}\rangle\langle {\alpha'}|,    
\end{equation}
in the basis set $\{|\alpha\rangle\}$ of the system of interest, into the following polynomial of products of powers of operators of a single bosonic mode $(\hat{a}, \hat{a}^\dagger)$, as follows:
\begin{equation}\label{Hmn2}
\begin{split}
\hat{H}_{\text{sbm}}=\sum_{m=0}^{k-1}\sum_{n=0}^{k-1}H_{nm}\hat{P}_{nm}.    
\end{split}
\end{equation}
where
\begin{equation}\label{sbh}
\hat{P}_{nm} \equiv \frac{1}{(k-1)!^{2}}\sqrt{\frac{m!}{n!}}(\hat{a}^\dagger)^n \hat{\Gamma}_k^{k-1} (\hat{a}^\dagger)^{k-1-m}.
\end{equation}
with
\begin{equation}\label{Gamma}
\hat{\Gamma}_k=((k-1)-\hat{N})\hat{a}, 
\end{equation}
where $\hat{N}=\hat{a}^\dagger\hat{a}$. Appendix~\ref{sec:DM} shows that $\hat{\Gamma}_k$ corresponds to the operator $\hat{S}^{\dagger}_+$ of the Dyson-Maleev transformation.\cite{Dyson1956_1,Dyson1956_2,Maleev1958,Dembiski1964}

To derive the mapping introduced by Eq.~(\ref{Hmn2}), we map the operators $|\alpha\rangle\langle \alpha'|$ introduced by Eq.~(\ref{Hmn}) into the corresponding transition operators $|m\rangle\langle n|$ in the basis of the 1-dimensional harmonic oscillator (HO), satisfying $\hat{a}|m\rangle=\sqrt{m}|m-1\rangle, \hat{a}^\dagger|m\rangle=\sqrt{m+1}|m+1\rangle$. 
We can verify that 
\begin{equation}
\label{eq:ok}
|0\rangle\langle k-1|=\frac{\hat{\Gamma}_k^{k-1}}{(k-1)!^{3/2}},
\end{equation} 
in the subspace of the first $k$ eigenstates of the HO (Appendix~\ref{sec:BD}).  Therefore, $\frac{\hat{\Gamma}_k^{k-1}}{(k-1)!^{3/2}}$ effectively acts as the transition operator $|0\rangle\langle k-1|$. As shown in Appendix~\ref{sec:BD}, the definition of $\hat{\Gamma}_k$ leads to a block diagonal representation of operators. For example, for $k=3$, we obtain:
\begin{equation*}\label{eq:occ7}
 \frac{\hat{\Gamma}_3^{2}}{2^{3/2}}=\left(
\begin{array}{ccc:ccccc}
0 & 0 & 1 & 0 & 0 & 0 & 0 & \dots\\
0 & 0 & 0 & 0 & 0 & 0 & 0 & \dots\\
0 & 0 & 0 & 0 & 0 & 0 & 0 & \dots\\
\hdashline
0 & 0 & 0 & 0 & 0 & \sqrt{10} & 0 & \dots\\
0 & 0 & 0 & 0 & 0 & 0 & 3\sqrt{15} & \dots\\
0 & 0 & 0 & 0 & 0 & 0 & 0 & \dots\\
0 & 0 & 0 & 0 & 0 & 0 & 0 & \dots \\
\dots & \dots & \dots & \dots & \dots & \dots & \dots & \ddots
\end{array} \right), 
\end{equation*}
showing that the matrix representation of $|0\rangle\langle 2|$ is indeed recovered from the top $3\times 3$ diagonal block. 

Next, substituting Eq.~(\ref{eq:ok}) into the expression of $\vert n \rangle \langle m \vert$, and considering that $\vert n \rangle = \frac{(\hat{a}^\dagger)^n}{\sqrt{n!}} \vert 0 \rangle$ and $\langle m \vert = \langle k-1 \vert (\hat{a}^\dagger)^{k-m-1} \sqrt{\frac{m!}{(k-1)!}}$, we obtain that any operator $\hat{P}_{nm} \equiv |n\rangle\langle m|$, with $n,m < k$, can be represented according to Eq.~(\ref{sbh}).

Note that Eq.~\eqref{Hmn2} is an operator of a single bosonic mode, which corresponds to a single $k$-qudit gate for the mapping of a $k\times k$ Hamiltonian. In particular, when $k=2$, Eq.~\eqref{Hmn2} provides the mapping of any $2\times 2$ hermitian operator into an operator of a single bosonic mode, allowing for construction of any bosonic 1-qubit gate with readily available superconducting devices. 

Appendix~\ref{sec:DM} describes the relationship between the SBM mapping and the established Dyson-Maleev (DM) and Holstein-Primakoff (HP) mappings, used to map spin operators into bosonic operators. The DM and HP mappings use one bosonic mode per spin site so they do not allow for the possibility of a single $k$-qudit gate. Furthermore, although both DM and HP mappings use bosonic operators, they are not able to construct the well-restricted bosonic Hamiltonian necessary for a quantum computing scheme. The DM mapping uses non-Hermitian bosonic operators which do not directly transfer to be unitary quantum gates upon exponentiation, while the operator square root term in the HP mapping is known to be hard to represent without a perturbative approach, which restricts implementation into quantum gates.

\section{Modular Quantum Circuits}\label{sec:SNAIL}
This section introduces a modular design of quantum circuits based on driven Superconducting Nonlinear Asymmetric Inductive eLements (SNAIL) with a capacitive shunt,\cite{frattini20173} parametrized according to SBM Hamiltonians. 

We begin by introducing the SBM mapping 
of $2\times 2$ hermitian matrices describing 1-qubit gates. 
The operator $\Gamma_k$ introduced by Eq.~\eqref{Gamma}, with $k=2$, is defined as follows:
\begin{equation}
\begin{split}
\hat{\Gamma}_2 &= (1 - \hat{a}^\dagger \hat{a}) \hat{a},\\
&= \hat{a} - \hat{a}^{\dagger} \hat{a}^2,
\end{split}
\end{equation}
so any $2 \times 2$ matrix can be written according to Eq.~\eqref{Hmn2}, as follows:
\begin{equation}
\hat{H}_{sbm} = \sum_{j,k=1}^2  H_{jk} \hat{P}_{j,k},
\end{equation}
where $\hat{P}_{1,2} = \hat{a}-\hat{a}^{\dagger} \hat{a}^2$, $\hat{P}_{2,2} = \hat{a}^{\dagger} \hat{a}$, $\hat{P}_{1,1} = 1-\hat{a}^{\dagger} \hat{a}$, and
$\hat{P}_{2,1} = \hat{a}^{\dagger}-(\hat{a}^{\dagger})^2  \hat{a}$.

Defining $H_{12}=R_{12} e^{i \phi_{12}}$ with real valued $R_{12}$ and $\phi_{12}$ and introducing the substitution $\hat{b} = \hat{a} e^{i \phi_{12}}$, we obtain: 
\begin{equation}
\begin{split}
\hat{H}_{sbm} 
&= H_{11} + (H_{22} - H_{11}) \hat{b}^{\dagger} \hat{b} + R_{12} (\hat{b}+\hat{b}^\dagger)-R_{12} \left(\hat{b}^{\dagger} \hat{b}^2 +(\hat{b}^{\dagger})^2  \hat{b} \right),\\
&=H_{11} + (H_{22} - H_{11}) \hat{b}^{\dagger} \hat{b} + R_{12} (\hat{b}+\hat{b}^\dagger)-R_{12} \left(\frac{(\hat{b}^{\dagger}+\hat{b})^3}{3} -(\hat{b}^\dagger+\hat{b})-\frac{\hat{b}^{{\dagger}^3}+\hat{b}^3}{3} \right),\\
&=H_{11} + \hbar \omega \hat{b}^{\dagger} \hat{b} + 2 R_{12} (\hat{b}+\hat{b}^\dagger) + g_3 (\hat{b}+\hat{b}^\dagger)^3 +g_3(\hat{b}^{\dagger^3}+\hat{b}^3),
\end{split}
\label{eq:haman12}
\end{equation}
where $\hbar \omega = H_{22} - H_{11}$, and $g_3=-\frac{R_{12}}{3}$. 

\begin{figure*}
\includegraphics[width = \columnwidth]{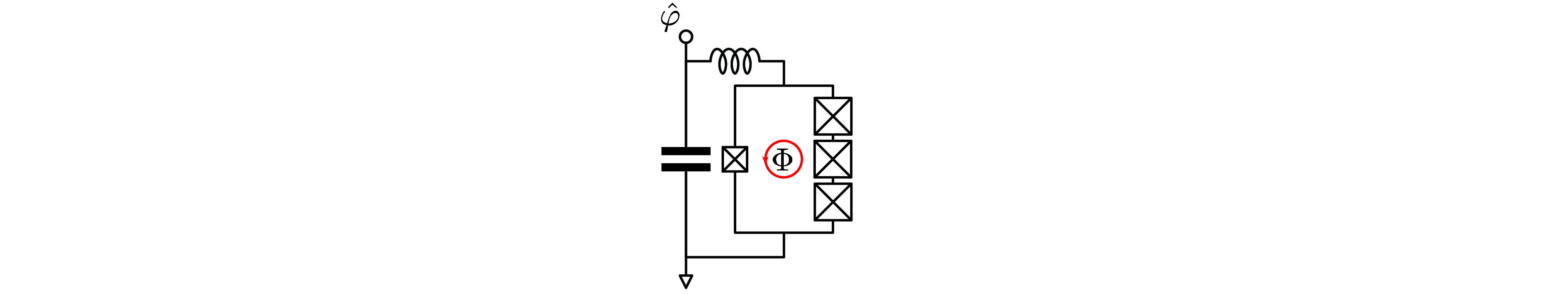}
\caption{Capacitively shunted SNAIL with series stray inductance. Cross-in-box symbol represents a Josephson tunnel junction, including an intrinsic shunt capacitance, while the circled arrow indicates a constant external flux $\Phi$ with tunable intensity.
The phase operator $\hat{\varphi}$ associated to the active node of the circuit (white-filled circle) is related to ladder operators via $\hat{\varphi} = \varphi_\mathrm{zpf}(\hat{b} + \hat{b}^\dagger)$.}
\label{fig:snail}
\end{figure*}

Considering that the Hamiltonian of a capacitively shunted SNAIL (Fig.~\ref{fig:snail}) is~\cite{frattini20173}
\begin{equation}
\hat{H}_{SPA}= \hbar \omega \hat{b}^\dagger \hat{b} + g_3 (\hat{b}+\hat{b}^\dagger)^3 + g_4 (\hat{b}+\hat{b}^\dagger)^4,
\label{eq:SPAH}
\end{equation}
we can readily identify the Hamiltonian $\hat{H}$, introduced by Eq.~(\ref{eq:haman12}), as the Hamiltonian of a linearly driven (displaced) SNAIL,
\begin{equation}
\begin{split}
\hat{H}_{sbm} 
&= H_{11} + 2 R_{12} (\hat{b}+\hat{b}^\dagger) + \hat{H}_{SPA} +g_3(\hat{b}^{\dagger^3}+\hat{b}^3),\\
\end{split}
\label{eq:haman4}
\end{equation}
with the fourth-order term turned off ($g_4=0$).

Note that the term $g_3(\hat{b}^{\dagger^3}+\hat{b}^3)$ in Eq.~\eqref{eq:haman4} can be produced by driving the SNAIL at a frequency $\omega_3 \approx 3\,\omega$. Indeed, a four-wave mixing interaction would be able to implement such term in a frame rotating at $\omega_3/3$ \cite{ChangPRX2020}. We want to emphasize that, despite the assumption of $g_4=0$, four-wave mixing can still be implemented by cascaded three-wave mixing processes \cite{frattini_optimizing}.

 More generally, a SNAIL can be substituted by an arbitrary flux-biased Josephson circuit~\cite{miano2023}, providing additional freedom for the choice of $\omega$ and $g_3$ coefficients in the Hamiltonian introduced by Eq.~\eqref{eq:SPAH}. Consequently, a wide range of combinations of the coefficients $H_{ij}$ can be engineered at the hardware level. We note that despite the generality of Eqs.~\eqref{eq:haman12} and \eqref{eq:haman4}, cases with $\omega\leq 0$ for a physical oscillator might be energetically unstable, which would impose limitations on the construction of arbitrary $2\times 2$ hermitian matrices. However, it is verified in Appendix~\ref{sec:onequbit} that all $R_z$ and $R_x$ gates can be implemented under this restriction. As these gates constitute a 1-qubit universal set, the hardware setting proposed in Fig.~\ref{fig:snail} can be used to construct arbitrary 1-qubit gates. 


 
\begin{figure}
  \centering
  \begin{subfigure}{1.\textwidth}
   \label{fig:subfiga}
    \centering
    \includegraphics[width=\linewidth]{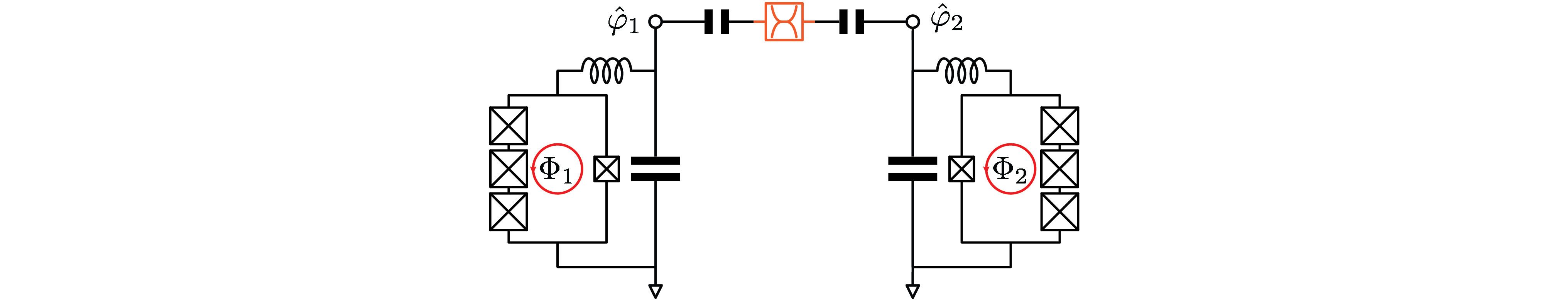}
    \caption{}
    \vspace{1cm}
  \end{subfigure}
  \begin{subfigure}{1.\textwidth}
    \centering
    \includegraphics[width=\linewidth]{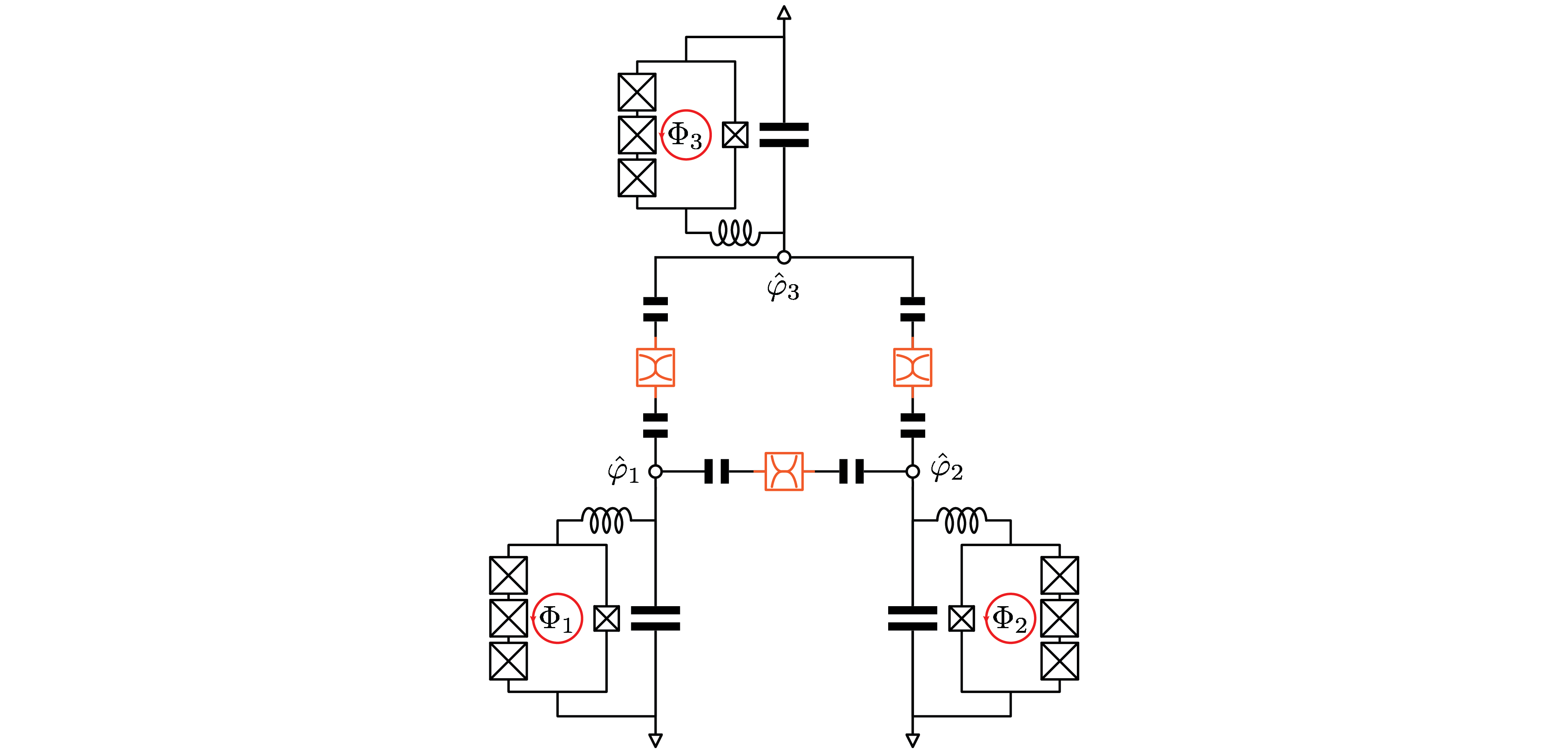}
    \caption{}
    \label{fig:subfigb}
  \end{subfigure}
  \caption{Modular assembly of capacitively shunted SNAILs for mapping (a) $4 \times 4$ and (b) $8 \times 8$ (bottom) hermitian matrices. The dynamics of the i-th capacitively shunted SNAIL is described by a phase operator $\hat{\varphi}_i = \varphi_{\mathrm{zpf}_i}(\hat{b}_i + \hat{b}_i^\dagger)$. The SNAILs are coupled via nearly-quartic elements, represented as distorted cross-in-box symbols (in orange). Consequently, the eigenmodes can be assumed to be the same of the uncoupled system.}
  \label{fig:snail23}
\end{figure}

 To establish a universal set of quantum gates, a 2-qubit entangling gate ({\em e.g.}, a controlled-Z gate) is required. This requirement can be fulfilled by a modular design of driven SNAIL circuits nonlinearly coupled by nearly-quartic elements, effectively described by a $4\times 4$ Hamiltonian, as shown in Fig.~\ref{fig:snail23}(a).

 A nearly-quartic element can be implemented, for instance, by a SNAIL designed with an unusual combination of Josephson junctions or a dc-SQUID.\cite{ye2021engineering}
 More in general, any superconducting two-terminal circuit whose potential energy function $U$ can be approximated as,
 \begin{equation}
 \label{eq:quartic_potential}
    U(\varphi) \approx \frac{a}{4!}(\varphi-\varphi_0)^4 + O((\varphi-\varphi_0)^5),
 \end{equation}
 can implement such nearly-quartic element. In Eq.~(\ref{eq:quartic_potential}), $\varphi$ is the phase difference across the terminals of the superconducting circuit implementing the potential energy $U$ and $a=\left.\frac{d^4U}{d\varphi^4}\right|_{\varphi_0}$ is the fourth-order Taylor expansion coefficient of the function $U$, evaluated at the point $\varphi_0$ which minimizes $U$. While it is possible to implement the potential energy in Eq.~(\ref{eq:quartic_potential}) exactly,~\cite{ye2021engineering} in practice, any two-terminal circuit including one or more Josephson tunnel junctions is shunted by an intrinsic capacitance that introduces a weak linear coupling between the two terminals. Such linear capacitive coupling arises from the intrinsic capacitance of the Josephson tunnel junctions, and can be neglected when the fourth order nonlinearity implemented by $U$ is the dominant coupling mechanism between the two terminals ({\em i.e.}, the "nearly-quartic" coupling limit).
 
 A nearly-quartic element can be used to implement ultra-strong cross-Kerr couplings~\cite{ye2021engineering} between photonic modes described by the interaction Hamiltonian,
 \begin{equation}
 \hat{H}_{\text{cross-Kerr}}=\chi\hat{b}_1^\dagger\hat{b}_1\hat{b}_2^\dagger\hat{b}_2, 
 \end{equation}
 where $\hat{b}_1$ and $\hat{b}_2$ are the annihilation operators of the two coupled photonic modes. The nearly-pure and ultra-strong cross-Kerr coupling can enable the construction of many photonic 2-qubits gates, including a controlled-Z gate as shown in Appendix~\ref{sec:quarton}. Therefore, with a combination of one-qubit gates and the two-qubit gate, constructed as quartic-connected SNAILs, it is possible to map any physical Hamiltonian into a modular cQED processor. 
 
 
 Design of a multiple-qubit entangling gate is realized by the circuit of multiple SNAILs coupled with nearly-quartic elements. As an example, Fig.~\ref{fig:snail23}(b) illustrates the circuit that effectively maps the $8\times 8$ Hamiltonian corresponding to a 3-qubit entangling gate. Alternatively, bilinear couplings can also be established by beams splitters,\cite{Zhou2022,Chapman2023} as previously investigated for transmons.\cite{gao2018programmable,9371955}
 The $4 \times 4$ and $8 \times 8$ circuits in Fig.~\ref{fig:snail23} can be generalized as well to include arbitrary flux-biased Josephson circuits as a replacement for the SNAILs and the nearly-quartic couplers.

\section{Dynamics of Charge and Energy Transfer}\label{sec:models}
\begin{figure*}
\includegraphics[scale=0.14]{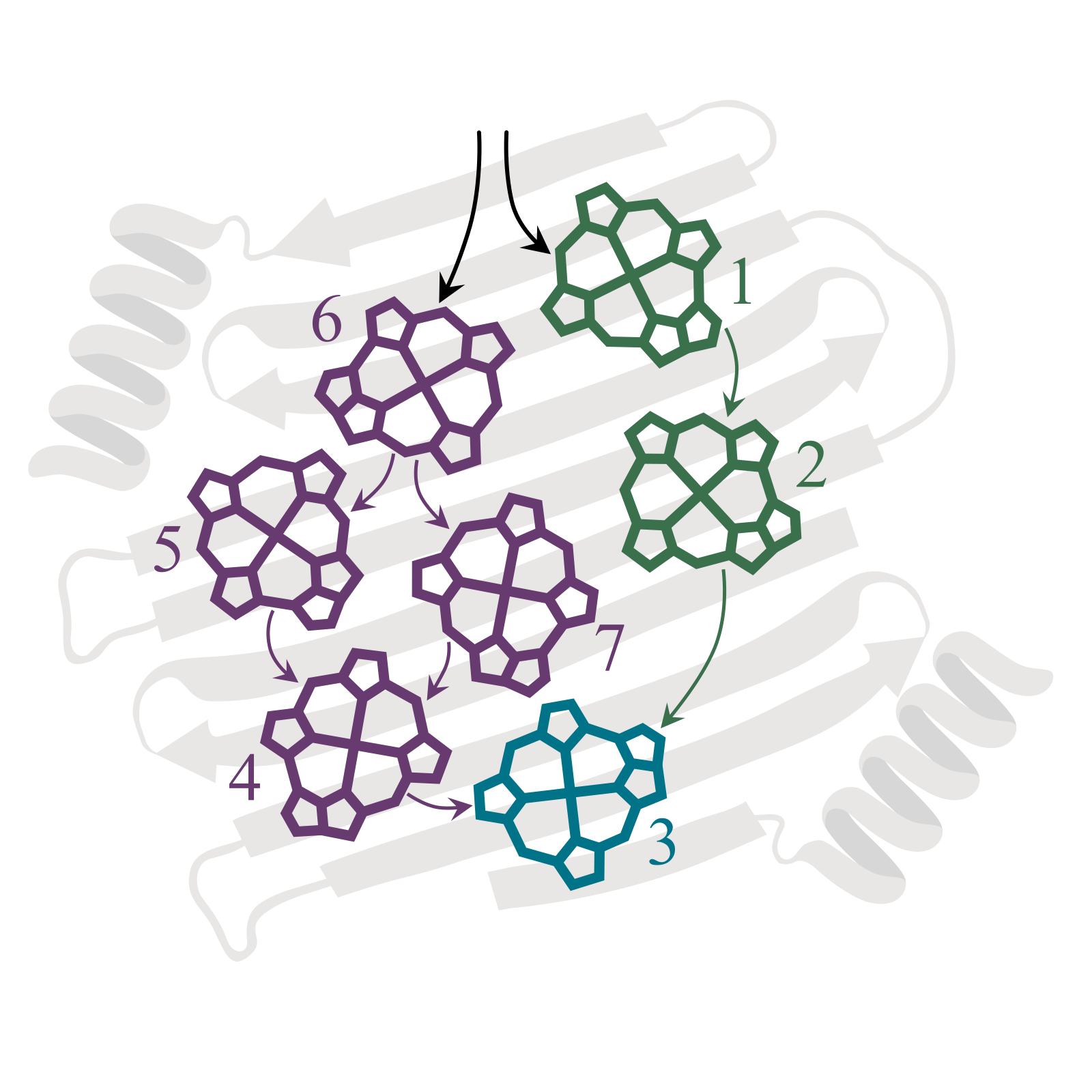}
\includegraphics[scale=0.35]{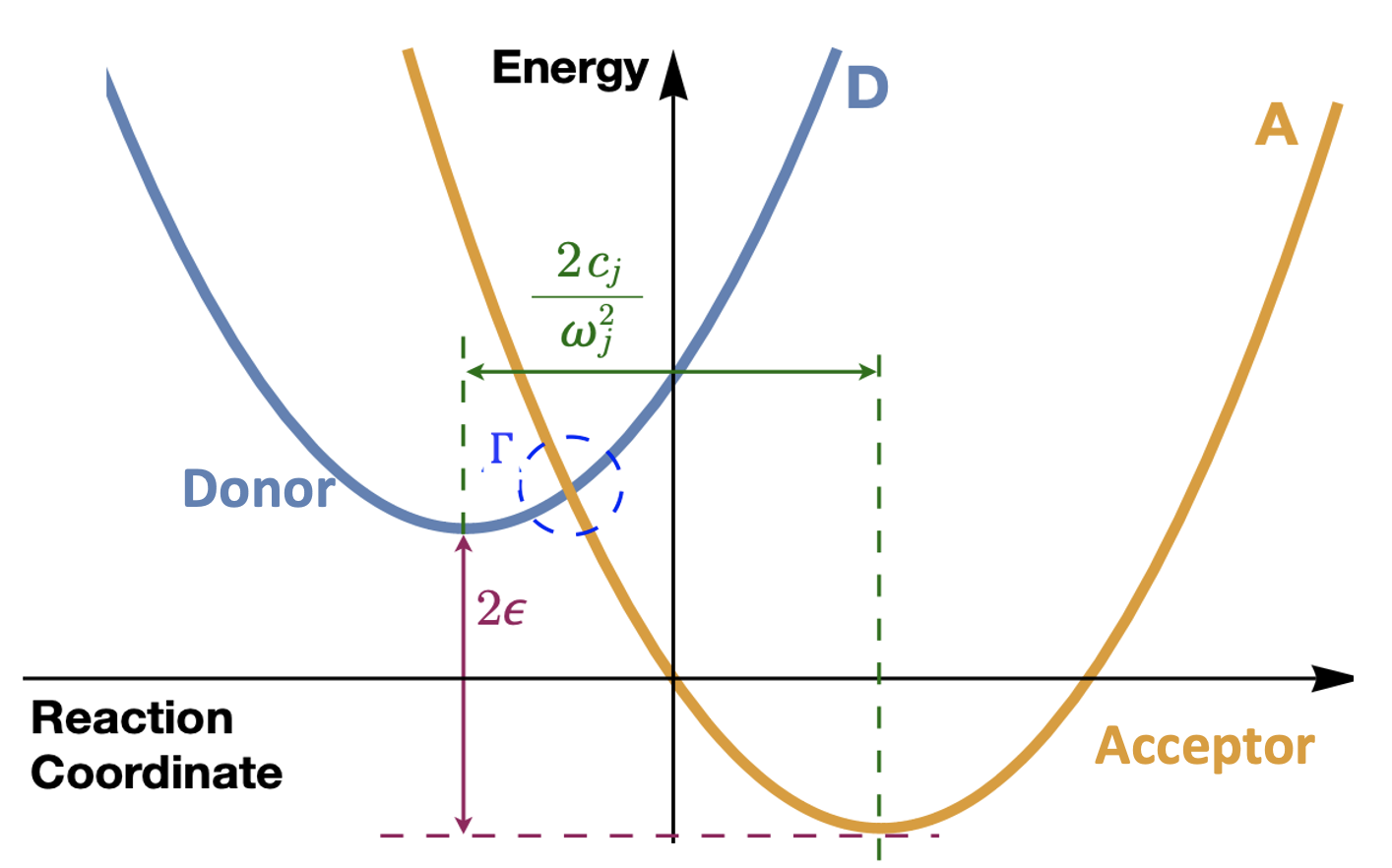}
\caption{Schematic representation of model systems for quantum dynamics simulations of energy transfer (up) and electron transfer (down).}
\label{fig:2fmo}
\end{figure*}

A variety of important dynamical processes in molecular systems of chemical, biological and technological importance involve electronic energy and charge transfer. The simulation of the inherently quantum-mechanical electronic dynamics underlying these processes is a subject of great interest. In this section, we illustrate the SBM mapping based on the SNAIL circuit as applied to quantum dynamics simulations of energy and charge transfer in model systems, including a four-level system describing energy transfer in the FMO light-harvesting complex, and a spin-boson model that describes charge transfer in the presence of dissipation, schematically represented in Fig.~\ref{fig:2fmo}. 

\subsection{Two-level system (TLS)}
The simplest model of energy or charge transfer is given by the $2 \times 2$ donor-acceptor Hamiltonian,~\cite{Tang2015} 
\begin{equation}
\begin{split}
\hat{H}_{TLS}=\begin{pmatrix}
-\epsilon&\Delta\\
\Delta&\epsilon\\
\end{pmatrix},
\end{split} 
\label{eq:TLSham}
\end{equation}  
describing two coupled electronic states, with $\epsilon=50~\text{cm}^{-1}$ and $\Delta=20~\text{cm}^{-1}$ for a typical charge transfer process in molecules. To map the $2\times 2$ Hamiltonian into the circuit of Fig.~\ref{fig:snail}, the SNAIL parameters are obtained according to Eq.~\eqref{eq:haman4}, with oscillator frequency $\omega=100~\text{cm}^{-1}$, linear displacement $R_{12}=50~\text{cm}^{-1}$ and third-order coupling $g_3=-16.7~\text{cm}^{-1}$. With these parameters, the right hand side of Eq.~\eqref{eq:haman4} is programmed on a classical computer and numerically exponentiated to obtain the corresponding propagator for dynamics simulations. Fig.~\ref{fig:sb_matvec}a shows the simulation results for the time-dependent population of the donor state. The exact agreement with benchmark calculations obtained by numerically integrating the Schr\"{o}dinger equation demonstrates the SBM mapping and the proposed SNAIL-based circuit. 

\subsection{Fenna–Matthews–Olson Complex }
Energy transfer through the chlorophyll pigments of the Fenna–Matthews–Olson (FMO) complex (Fig.~\ref{fig:2fmo}a) corresponds to exciton transfer across chromophore sites. The excitons are modeled as hard-core bosons~\cite{higgins2017quantum,hu2018connecting,hu2018double}, according to the Frenkel exciton Hamiltonian,\cite{abramavicius2011exciton}
\begin{equation}
    H={{E}_{j}}\sum\limits_{j}^{{}}{\sigma_{j}^{+}\sigma_{j}^{-}+{{J}_{jk}}\sum\limits_{j,k}^{{}}{(\sigma_{j}^{+}\sigma_{k}^{-}+\sigma_{k}^{+}\sigma_{j}^{-})}},
\end{equation}
where ${{\sigma_j}^{+}}$ and ${{\sigma_j }^{-}}$ are the Pauli-raising operator and lowering operators,
corresponding to the creation and annihilation of an excitation in chromophore $j$, with commutation rules $[{{\sigma_j}^{-}},{{\sigma_k}^{+}}]=\delta_{jk}(1-2~{{\sigma_j}^{+}}{{\sigma_k}^{-}})$.

The Hamiltonian can be written in the basis of chromophore occupation number. We consider the energy transfer through sites 1-4  (Fig.~\ref{fig:2fmo}a), as described by the following $4 \times 4$ Hamiltonian matrix:~\cite{Schulze2016}
\begin{equation}
\hat{H}_{FMO}=\begin{pmatrix}
310.0 & -97.9 & 5.5 & -5.8 \\
-97.9 & 230.0 & 30.1 & 7.3 \\
5.5 & 30.1 & 0.0 & -58.8\\
-5.8 & 7.3 & -58.8 & 180.0
\end{pmatrix},
\label{eq:FMOham}
\end{equation}
with parameters in cm$^{-1}$. Diagonal terms correspond to the energies of the chromophore while off-diagonal terms are the couplings between them. 

To parametrize the superconducting circuit for dynamics simulations, with an integration time-step $\tau$, we obtain the propagator $\hat{U}_{FMO}=e^{-i\tau\hat{H}_{FMO}/\hbar}$ as a $4\times 4$ unitary matrix. This 2-qubit gate is then transpiled in terms of SNAILs parametrized according to the set of elementary gates including 1-qubit rotations and controlled-Z gates. Note that
we are able to convert the Pauli operators into single boson operators based on the SBM mapping, offering advantages
over conventional bosonization methods such as the Holstein–Primakoff~\cite{Holstein1940}, or the Dyson–Maleev transformation~\cite{Dyson1956_1,Dyson1956_2,Maleev1958,Dembiski1964}
(Appendix \ref{sec:DM}). Analogous implementations could also be applied to model fermionic Hamiltonians commonly encountered in quantum chemistry, when converted into sums of tensor products of Pauli operators in conjunction with the Jordan-Wigner transformation and then mapped into bosonic gates. 

\begin{figure*}
\includegraphics[scale=0.6]{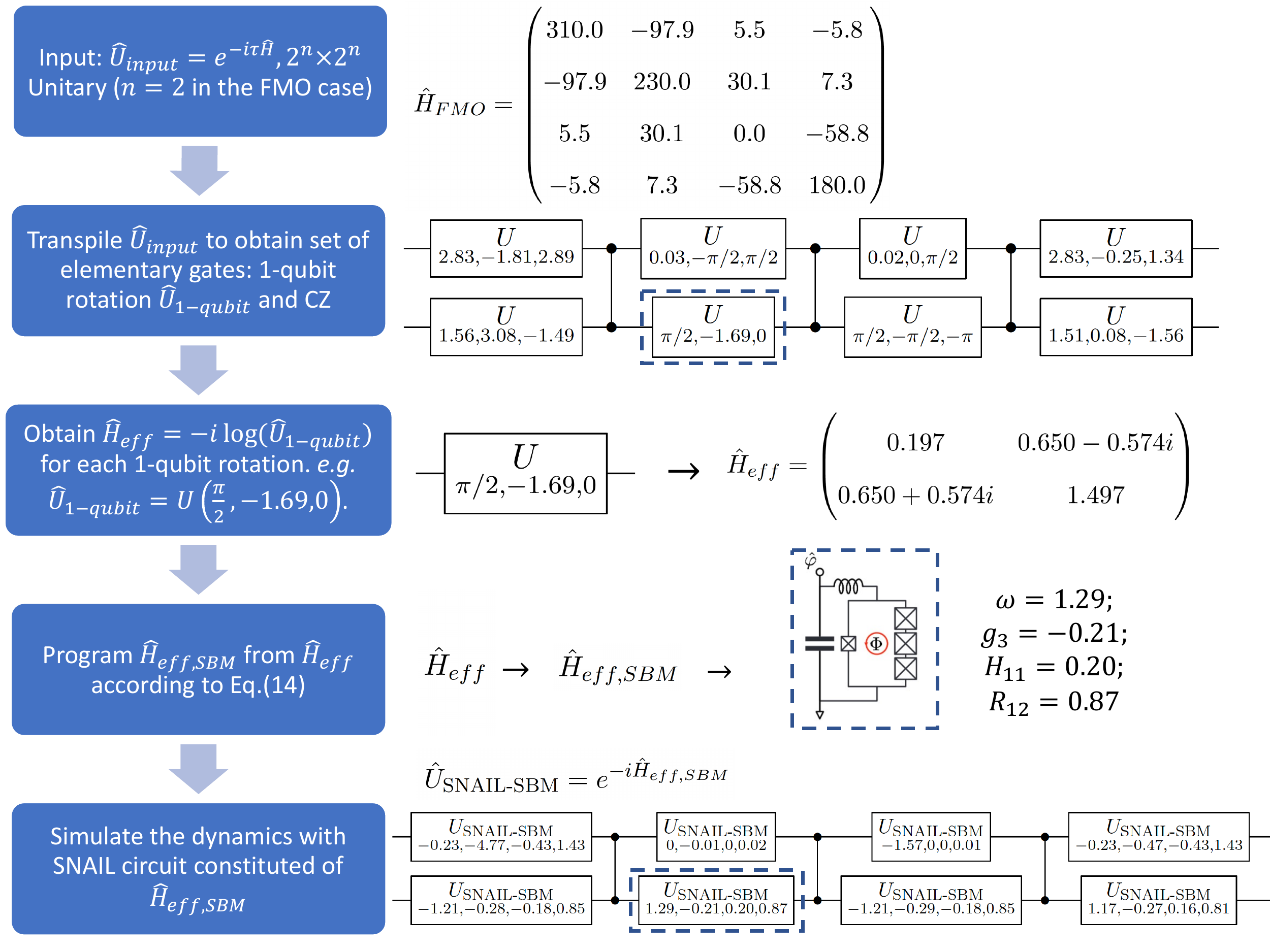}
\caption{Scheme for simulating the SBM-SNAIL circuit that propagates energy transfer in the FMO 4-site model. The three parameters in the first circuit diagram are rotation angles that define the 1-qubit rotation operations, following Ref.~\citenum{qiskit_2023}. The four parameters in the bottom circuit diagram are the SNAIL gate parameters in Eq.~\eqref{eq:haman4}. From left to right: oscillator frequency $\omega$, third-order coupling term $g_3$, constant term $H_{11}$ and half of displacement $R_{12}$. }
\label{fig:SNAIL_scheme}
\end{figure*}

To obtain the SNAIL parameters for a 1-qubit rotation $\hat{U}_{\text{1-qubit}}$, we compute the effective Hamiltonian $\hat{H}_{\text{eff}}=-i~\text{log}(\hat{U}_{\text{1-qubit}})$, then we map that Hamiltonian as $\hat{H}_{\text{eff,sbm}}$ according to Eq.~\eqref{eq:haman4}, and we obtain the corresponding rotation gate, as follows: $\hat{U}_{\text{eff,SBM}}=e^{-i\hat{H}_{\text{eff,sbm}}}$. The circuit is simulated by arranging the gates $\hat{U}_{\text{eff,SBM}}$ according to the transpiled circuit diagram, with CZ gates corresponding to two SNAILs coupled by a nearly-quartic element, as described in Sec.~\ref{sec:SNAIL}. Fig.~\ref{fig:SNAIL_scheme} shows a schematic representation of the resulting simulation.  

\begin{figure*}
\includegraphics[scale=0.8]{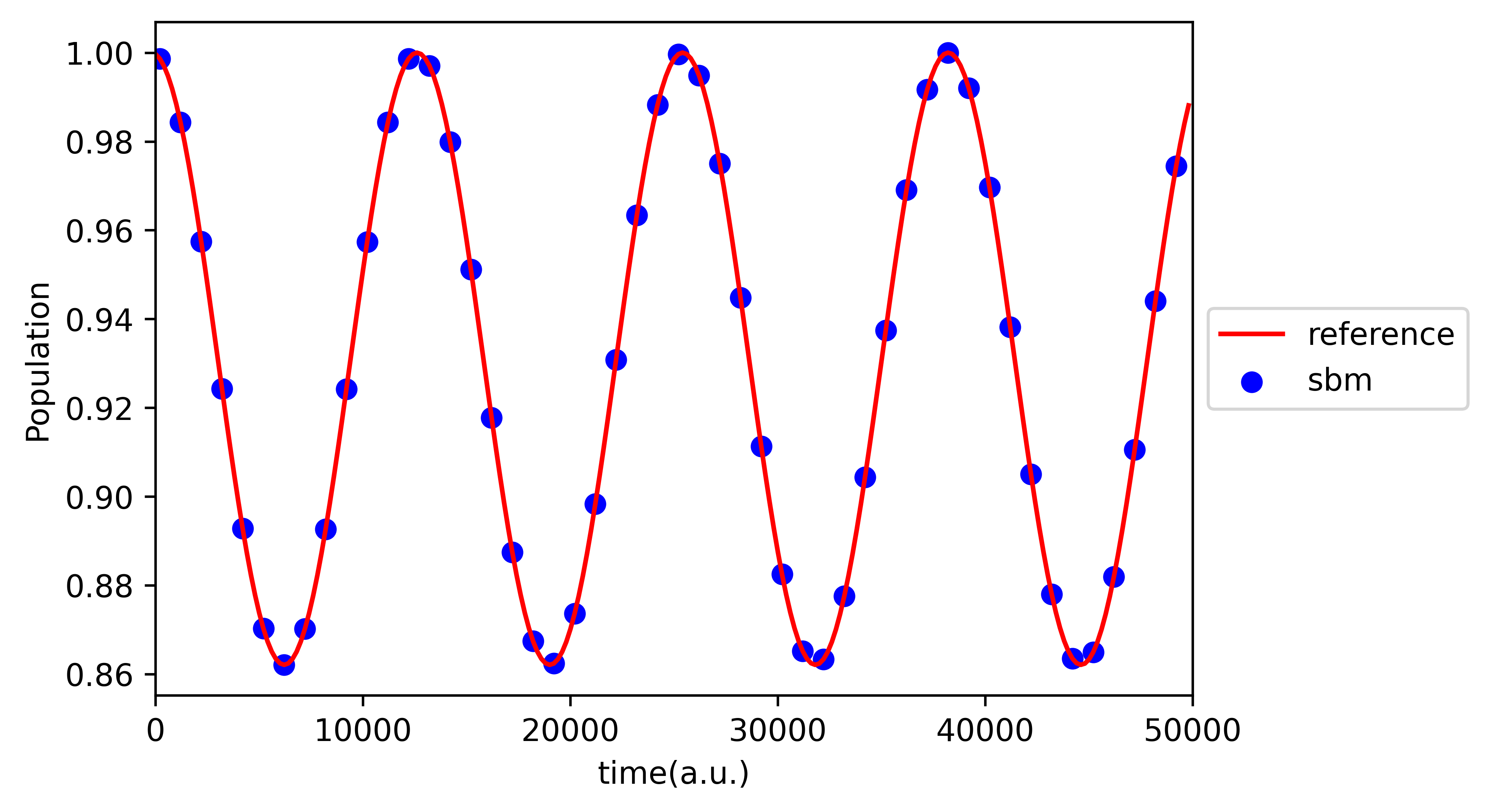}
\includegraphics[scale=0.8]{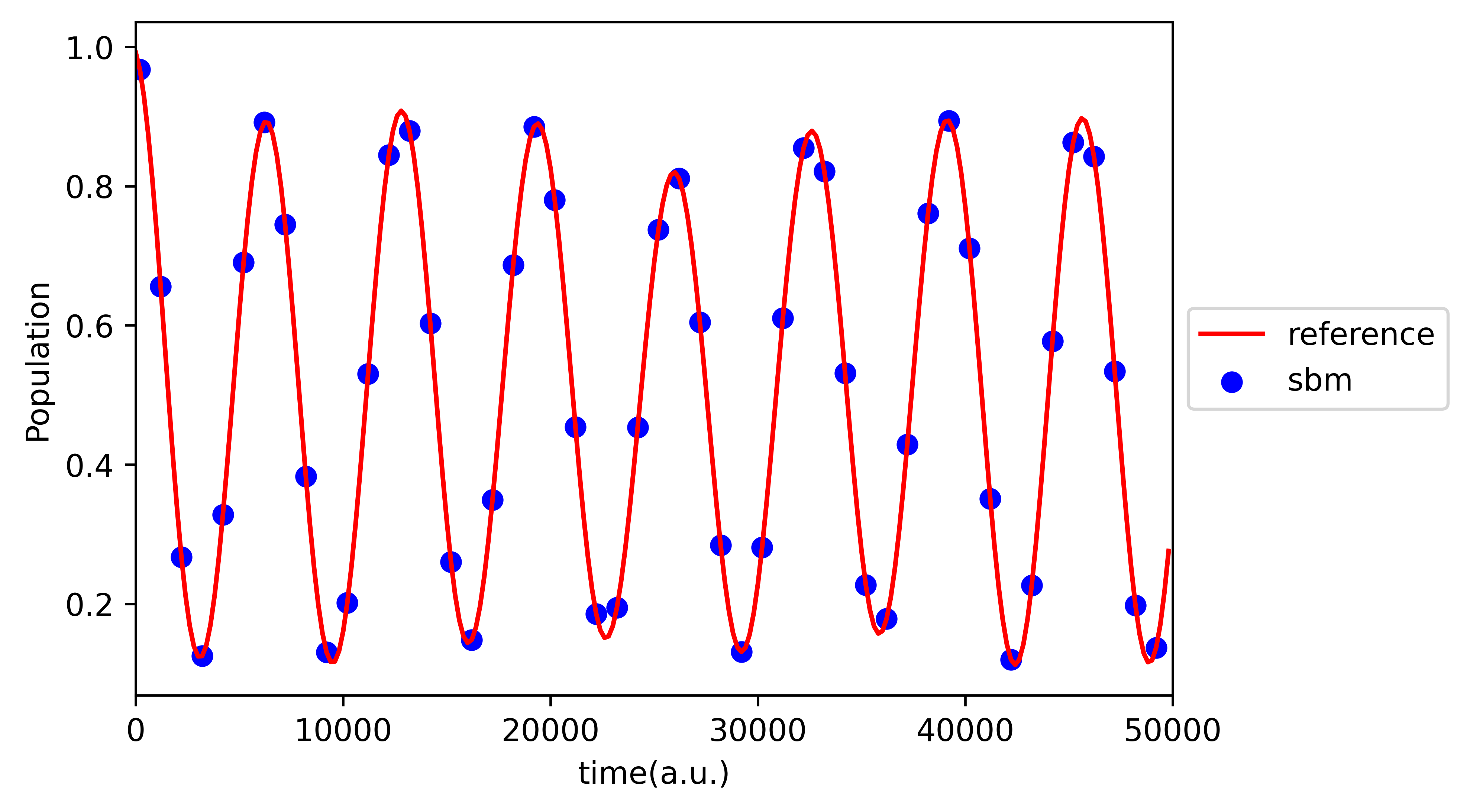}
\caption{Population dynamics for (a) spin-up state of the two-level system and (b) chlorophyll 1 of the 4-site FMO complex, obtained from single-boson-mapped Hamiltonians with simulated SNAIL circuit (blue dots), benchmarked with results obtained from directly integrating the Schr\"{o}dinger equation. (red lines).}
\label{fig:sb_matvec}
\end{figure*}

Fig.~\ref{fig:sb_matvec}b shows the results of simulations of the exciton dynamics for site 1, which is initially fully populated and gets depopulated according to the energy transfer process. The agreement between the results obtained with the SBM-mapped Hamiltonian and the reference calculations further demonstrates the capabilities of the SBM-SNAIL circuit design.

\subsection{Dynamics of Open Quantum Systems}
This section demonstrates the capabilities of the modular design of quantum circuits based on the SBM-mapping, as applied to dynamics simulations of open quantum systems.
We focus on the spin-boson model including two electronic states coupled to a bath of displaced harmonic oscillators, described in Appendix~\ref{sec: Popu only}, recently analyzed with
tensor-train thermo-field memory kernels for generalized quantum master equations.~\cite{Lyu2023} 

Our propagation scheme is based on the so-called population-only Liouville space superoperator $\mathcal{P}^{pop}(t)$ that satisfies the following equation:
\begin{equation}\label{vmatvec}
\hat{\sigma}^{\text{pop}}(t)=\mathcal{P}^{\text{pop}}(t)\hat{\sigma}^{\text{pop}}(0),
\end{equation}
where $\hat{\sigma}(t)=\text{Tr}_n[\hat{\rho}(t)]$ is the reduced density matrix for the electronic DOFs, with $\hat{\rho}(t)$ the density matrix for the full vibronic system. Here, $\hat{\sigma}^{\text{pop}}(t)=(\sigma_{00}(t),\sigma_{11}(t))^T$ includes only the diagonal elements of $\hat{\sigma}(t)$, necessary to describe the electronic population dynamics. The preparation of the super-operator $\mathcal{P}^{pop}(t)$ is described in Appendix~\ref{sec: Popu only}. 

We compare the elements of $\hat{\sigma}^{\text{pop}}(t)$ obtained according to Eq.~\eqref{vmatvec} with the corresponding time-dependent populations obtained according to the quantum computational scheme based on the SBM-mapping. To perform quantum computing simulations based on Eq.~\eqref{vmatvec}, we first transform $\mathcal{P}^{\text{pop}}(t)$ into a unitary matrix using the Sz.-Nagy dilation theorem\cite{nagy1970harmonic}, as follows:~\cite{hu2020quantum,levy2014dilation} 
\begin{equation}
\mathcal{U}_{\mathcal{P}^{\text{pop}}}(t)=\begin{pmatrix}
\mathcal{P}^{\text{pop}}(t)&\sqrt{I-\mathcal{P}^{{\text{pop}}}(t)\mathcal{P}^{\text{pop}^\dagger}(t)}\\
\sqrt{I-\mathcal{P}^{\text{pop}{^\dagger}}(t)\mathcal{P}^{\text{pop}}(t)}&-\mathcal{P}^{\text{pop}^\dagger}(t)\\
\end{pmatrix}.
\end{equation}
The vectorized $v_{\sigma(0)}$ is dilated by appending ancillary zero elements, as follows:
\begin{equation}
\hat{\sigma}^{\text{pop}}(0)=(\sigma_{00}(0),\sigma_{11}(0))^T\rightarrow \tilde{\sigma}^{\text{pop}}(0)=(\sigma_{00}(0),\sigma_{11}(0),0,0)^T.
\end{equation}
The dilated time-updated population-only density matrix is obtained, as follows:
\begin{equation}\label{diamatvec}
\tilde{\sigma}^{\text{pop}}(t)=\mathcal{U}_{\mathcal{P}^{\text{pop}}}(t)\tilde{\sigma}^{\text{pop}}(0).
\end{equation} 

\begin{figure*}
\includegraphics[scale=0.4]{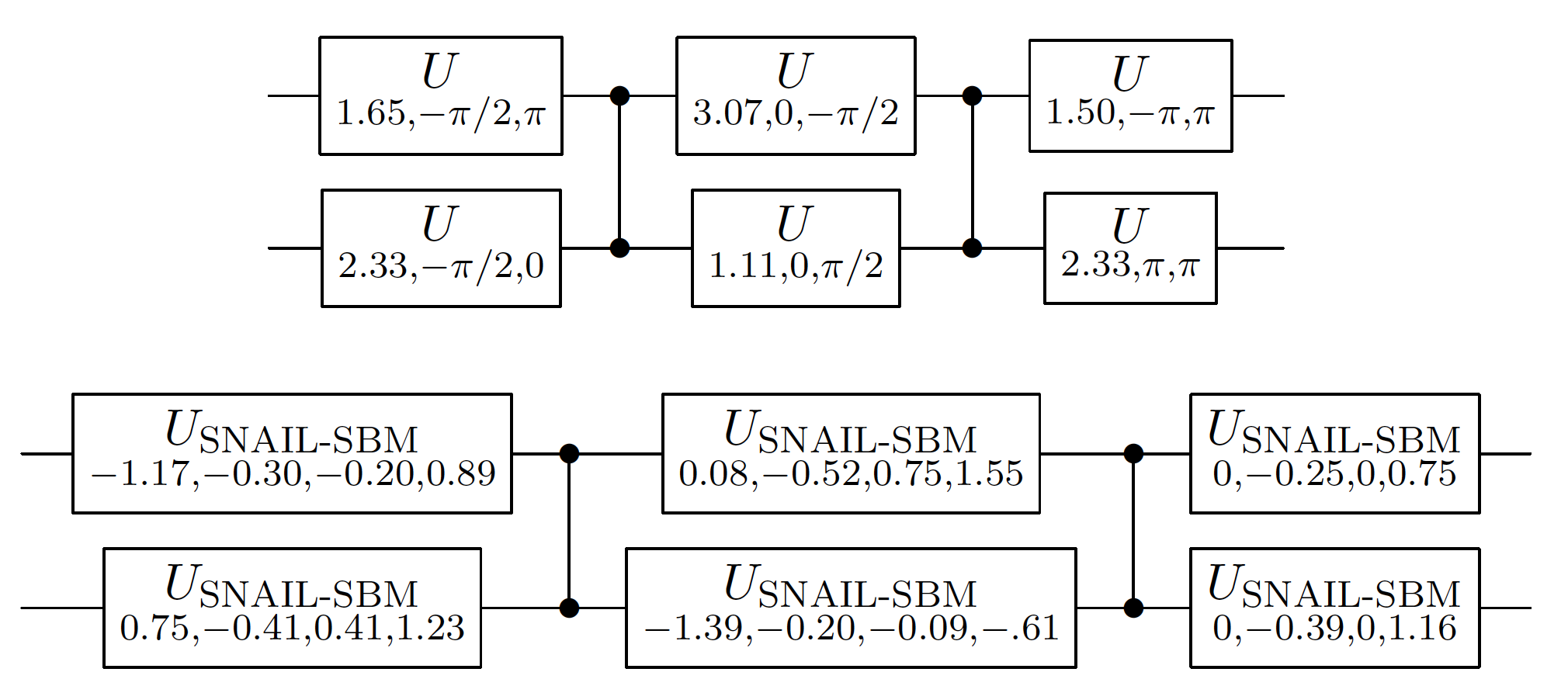}
\caption{Transpiled circuit (up) and SNAIL circuit (down) for $\cal{U}_{\cal{P}^{\text{pop}}}$$(t=1 a.u.)$.}
\label{fig:open_circuit}
\end{figure*}

The dilation scheme thus provides the unitary matrix $\mathcal{U}_{\mathcal{P}^{\text{pop}}}(t)$ governing the time-evolution of $\tilde{\sigma}^{\text{pop}}(t)$, the first two digits of which agree with those of $\hat{\sigma}^{\text{pop}}(t)$. Therefore, Eqs.~\eqref{diamatvec} and~\eqref{vmatvec} describe the same dynamics, with Eq.~\eqref{diamatvec} allowing for simulations on a quantum device. 
For the spin-boson model of interest $\mathcal{U}_{\mathcal{P}^{\text{pop}}}(t)$ is a $4\times 4$ unitary matrix, corresponding to a 2-qubit gate. Therefore, the SBM-SNAIL circuit is analogous to that of the FMO 4-site model. The simulation of the circuit thus follows the scheme of Fig.~\ref{fig:SNAIL_scheme}. The transpiled circuit and the corresponding SNAIL gate parameters for $\mathcal{U}_{\mathcal{P}^{\text{pop}}}(t=1 a.u.)$ are given in Fig.~\ref{fig:open_circuit}. 

\begin{figure*}
\includegraphics[scale=0.8]{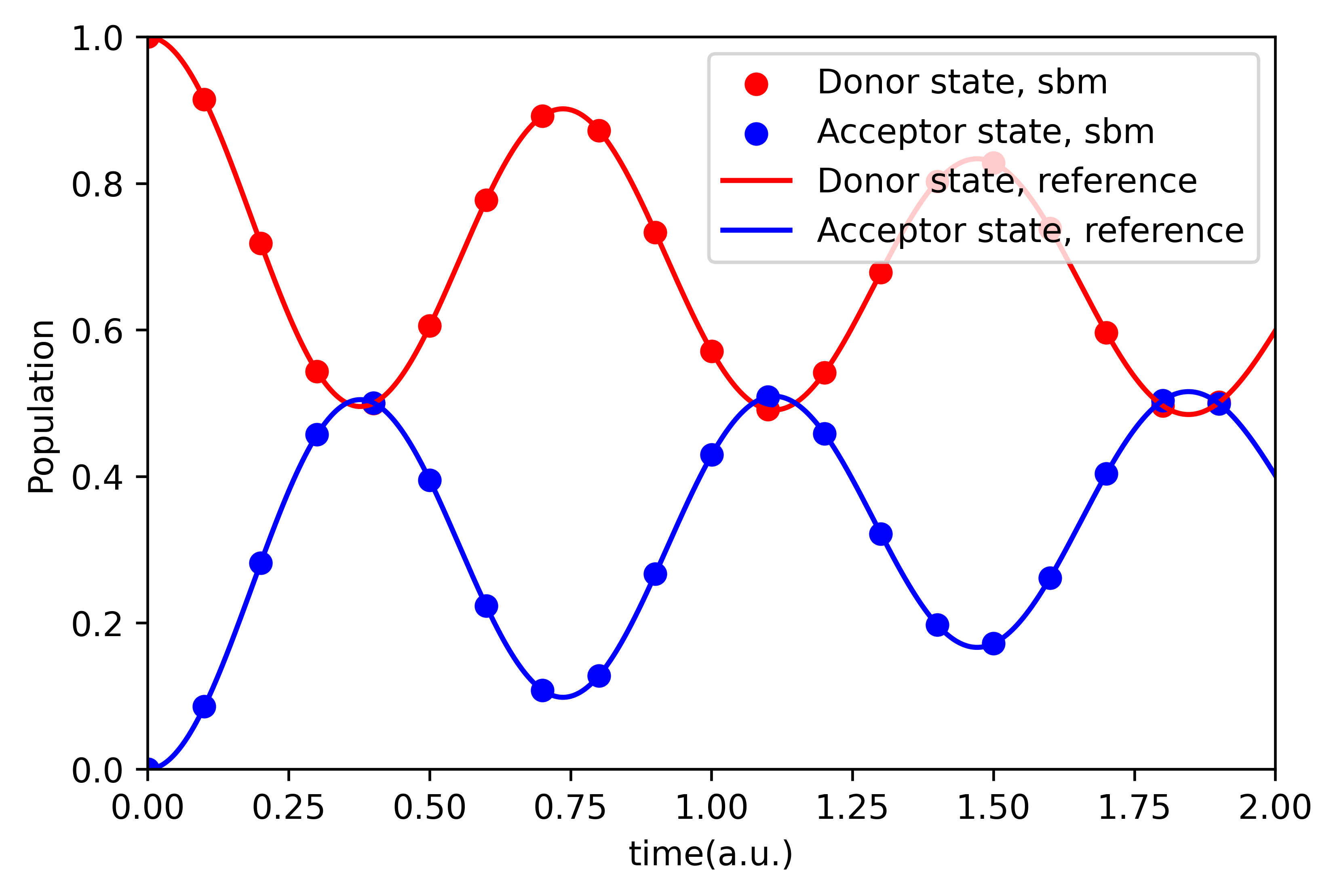}
\caption{Time dependent population of the electronic states, corresponding to the spin-boson model, obtained from the SBM-mapped $\mathcal{U}_{\mathcal{P}^{\text{pop}}}(t)$ matrices with simulated SNAIL circuit implementation (red and blue dots), benchmarked with the dynamics obtained directly from the original $\mathcal{P}^{\text{pop}}(t)$ matrices (red and blue lines).}
\label{fig:sbm_sb}
\end{figure*}

Figure~\ref{fig:sbm_sb} shows the comparison of time-dependent populations for the two electronic states corresponding to the spin-boson model, as described by elements of $\tilde{\sigma}^{\text{pop}}_{\text{sbm}}(t)$ obtained with the SBM-mapping with SNAIL circuit scheme, and the corresponding populations $\tilde{\sigma}^{\text{pop}}(t)$ obtained directly with Eq.~\eqref{diamatvec}, with initial condition $\tilde{\sigma}^{\text{pop}}(0)=(1,0,0,0)^T$. The excellent agreement demonstrates the capabilities of the SBM mapping as applied to a model of electron transfer with dissipation due to coupling to a surrounding environment. 

\section{Concluding Remarks}\label{sec:conclusion}
We have introduced a general method to map the Hamiltonian of molecular systems into the Hamiltonian of quantum circuits for cQED simulations. 
Additionally, we have identified the non-linear bosonic components that need to be assembled for a modular implementation of the corresponding circuit Hamiltonians. 
We have illustrated the SBM mapping, in conjunction with SNAIL circuits, as applied to simulations of energy transfer in the photosynthetic FMO model system, and charge transfer in donor-acceptor systems coupled to a dissipative environment. 

Beyond the modular design based on SNAILs, we have shown that the SBM mapping allows for implementation of Hamiltonians in the basis of qudits ({\em i.e.}, Eq.\eqref{Hmn2}, with $N>2$), corresponding to continuous-variable (CV) modes represented as N-dimensional discrete-variable (DV) states. 
For circuits with multiple qudits, the cross-Kerr Hamiltonian may also be generalized to perform a qudit controlled-Z gate allowing for construction of a universal set of gates for simulations on bosonic devices. The hardware efficiency of a qudit-based cQED can significantly reduce the circuit depthand simplify the experimental setup, offering a promising strategy for simulations of chemical systems. 

\section{Acknowledgements}
The authors acknowledge support from the NSF grant 2124511 [CCI Phase I: NSF Center for Quantum Dynamics on Modular Quantum Devices (CQD-MQD)]. We thank Ellen Mulvihill for helpful discussions and for preparing Fig.~\ref{fig:2fmo}. N.L. thanks Micheline B. Soley and Paul Bergold for stimulating discussions. 

\appendix

\section{Dyson-Maleev and Holstein-Primakoff Maps}\label{sec:DM}

Dyson and Maleev introduced a transformation~\cite{Dyson1956_1,Dyson1956_2,Maleev1958,Dembiski1964} to represent spin operators in terms of bosonic operators according to the ladder operators,
\begin{equation}
\begin{split}
\hat{S}_+ &= \hat{a}^{\dagger}\left[2s-\hat{N}\right],\\
\hat{S}_- &= \hat{a},\\
\hat{S}_z &= \hat{N}-s,
\end{split}
\label{DM_lower}
\end{equation}
with $\hat{N}=\hat{a}^\dagger \hat{a}$, and $[\hat{a}, \hat{a}^\dagger]=1$. 
Notice that $\hat{S}^{\dagger}_- \neq \hat{S}_+$, so the ladder operators are not Hermitian conjugates of each other and thus the transformation is not unitary. Nevertheless, Eqs. (\ref{DM_lower}) satisfy the Lie algebra of the original spin operators,
\begin{equation}
\hat{S}_\pm= \hat{S}_x \pm i\hat{S}_y,
\end{equation}
\begin{equation}
\left[\hat{S}_{+},\hat{S}_{-}\right] = 2 \hat{S}_z,
\end{equation}
and
\begin{equation}
\left[\hat{S}_{i},\hat{S}_{j}\right] = i\sum^{3}_{k=1}\varepsilon_{ijk}\hat{S}_{k},
\end{equation}
where $\varepsilon_{ijk}$ is the levi-citta symbol and $i,j,k \in \{x,y,z\}$.
If we replace the magnitude of the spin, $s$, by the number $(k-1)/2$ in Eq. (\ref{DM_lower}), we obtain:
\begin{equation}
\hat{S}^{\dagger}_+ =((k-1)-\hat{N})\hat{a}.
\label{eq:eq30}
\end{equation}
Comparing Eq.~(\ref{eq:eq30}) and Eq.~(\ref{Gamma}),  we see that $\hat{S}^{\dagger}_+$ is identical to the operator $\hat{\Gamma}_k$. The other operator that we use in the SBM mapping is $\hat{a}$, which is in turn the operator $\hat{S}_-$ introduced by Eq. (\ref{DM_lower}). Therefore, our SBM mapping implements the raising and lowering operators of the Dyson-Maleev transformation. The major difference between the SBM mapping and DM transformation lies in the fact that the DM mapping
replaces the operators $\hat{S}_+$ and $\hat{S}_-$, according to Eqs.~(\ref{DM_lower}), and therefore generates a Hamiltonian in terms of $\hat{a}$ and $\hat{a}^\dagger$ that is not Hermitian. On the other hand, the SBM mapping preserves the Hermitian property by using the $\hat{P}_{nm}$ operators to map the matrix elements of the Hamiltonian so the full matrix representation is automatically preserved. 

Similar to the Dyson-Maleev transformation,\cite{Dyson1956_1,Dyson1956_2,Maleev1958,Dembiski1964} the Holstein-Primakoff transformation~\cite{Holstein1940} maps the spin operators for a spin-$s$ particle to bosonic operators, as follows:
\begin{equation}
\label{eq:lhp}
\begin{split}
\hat{S}_+ &=\hat{a}^\dagger \sqrt{2s-\hat{N}},\\
\hat{S}_- &=\sqrt{2s-\hat{N}}\hat{a},\\
\hat{S}_z &=\hat{N}-s.
\end{split}
\end{equation}


Comparing Eq.~\eqref{Gamma} to Eqs.~(\ref{eq:lhp}), we see that here $\hat{S}^\dagger_+=\hat{S}_-$ but differs from the operator $\hat{\Gamma}_k$ by a factor of $\sqrt{2s-\hat{N}}$. Unfortunately, the square root of the number operator is challenging to implement without relying upon a perturbative expansion, which is only accurate when $s$ is sufficiently large. In contrast, the SBM mapping is generally applicable.

\section{Block-Diagonality}
\label{sec:BD}

In this section, we prove that the right-hand side (rhs) of Eq.~\eqref{Hmn2} is block-diagonal, ensuring that the physical space of states $\vert j \rangle$ with $j<k$ remains decoupled from the unphysical space of states $\vert j \rangle$ with $j \geq k$. Specifically, we show that the rhs of Eq.~\eqref{Hmn2} has the following block-diagonal form:
\begin{equation}\label{eq:Hmnmat}
\hat{H}_{sbm}=\left(
\begin{array}{cccc:cccc}
H_{0,0} & H_{0,1} & \cdots & H_{0,k-1} & 0 & 0 & 0 & \dots\\
H_{1,0} & H_{1,1} & \cdots & H_{1,k-1} & 0 & 0 & 0 & \dots\\
\vdots & \vdots & \ddots & \vdots & 0 & 0 & 0 & \dots\\
H_{k-1,0} &H_{k-1,1} & \cdots & H_{k-1,k-1} & 0 & 0 & 0 &\dots\\
\hdashline
0 & 0 & 0 & 0 & \text{X} & \text{X} & \text{X} & \dots\\
0 & 0 & 0 & 0 & \text{X} & \text{X} & \text{X} & \dots\\
0 & 0 & 0 & 0 & \text{X} & \text{X} & \text{X} & \dots\\

\vdots & \vdots & \vdots & \vdots & \vdots & \vdots & \vdots & \ddots
\end{array} \right) 
\end{equation}
To achieve this, we show that $\hat{P}_{nm}$, introduced by Eq.~\eqref{sbh}, has the block-diagonal form,
\begin{equation}\label{eq:Gamma_mn_mat}
\hat{P}_{nm}=\left(
\begin{array}{cccc:cccc}
0 & 0 & \cdots & 0 & 0 & 0 & 0 & \dots\\
0 & 1 & \cdots & 0 & 0 & 0 & 0 & \dots\\
\vdots & \vdots & \ddots & \vdots & 0 & 0 & 0 & \dots\\
0 & 0 & \cdots & 0 & 0 & 0 & 0 &\dots\\
\hdashline
0 & 0 & 0 & 0 & \text{X} & \text{X} & \text{X} & \dots\\
0 & 0 & 0 & 0 & \text{X} & \text{X} & \text{X} & \dots\\
0 & 0 & 0 & 0 & \text{X} & \text{X} & \text{X} & \dots\\

\vdots & \vdots & \vdots & \vdots & \vdots & \vdots & \vdots & \ddots
\end{array} \right),     
\end{equation}
where only the nm-$th$ element is equal to 1. Substituting Eq.~\eqref{eq:Gamma_mn_mat} into Eq.~\eqref{Hmn2} yields the matrix form in Eq.~\eqref{eq:Hmnmat}.

First, we show that $\langle n \vert \hat{\Gamma}_k^{k-1} |j\rangle = 0$, for all $j,n <k$, unless $j=k-1$ and $n=0$. Considering that
\begin{equation}\label{eq:gam_a}
\begin{split}
\hat{\Gamma}_k|j\rangle &=((k-1)\hat{I}-\hat{N})\hat{a}|j\rangle \\
&=(k-j)\sqrt{j}|j-1\rangle,\\
\hat{\Gamma}_k^2 |j\rangle, 
&=(k-j)(k-(j-1)) \sqrt{j (j-1)}|j-2\rangle,
\end{split}
\end{equation}
we obtain
\begin{equation}\label{eq:gam_a}
\begin{split}
\hat{\Gamma}_k^{l-1} |j\rangle 
&=(k-j)(k-(j-1)) \cdots (k-(j-(l-2))) \\
& \times \sqrt{j (j-1) \cdots (j-(l-2))}|j-(l-1)\rangle.
\end{split}
\end{equation}
So, $\langle n \vert \hat{\Gamma}_k^{k-1} |j\rangle = 0$, unless $j-(k-1)=n$, a condition that can only be fulfilled for $j,n < k$ when $j=k-1$ and $n=0$, for which
\begin{equation}\label{eq:Gam_0}
\hat{\Gamma}_k^{k-1} | k-1 \rangle = (k-1)!^{3/2} \vert 0 \rangle.
\end{equation}


Now we prove Eq.~\eqref{eq:Gamma_mn_mat} by showing that $\langle j \vert \hat{P}_{nm} \vert l \rangle=\delta_{jn} \delta_{lm}$. We start by showing that $\langle n \vert \hat{P}_{nm} \vert m \rangle=1$, as follows:
\begin{equation}\label{eq:mnth}
\begin{split}
\langle n \vert \hat{P}_{nm} \vert m \rangle&=\langle n|\frac{1}{(k-1)!^2}\sqrt{\frac{m!}{n!}}(\hat{a}^\dagger)^n \hat{\Gamma}_k^{k-1} (\hat{a}^\dagger)^{k-1-m}|m\rangle\\
&=\langle 0|\frac{1}{(k-1)!^2}\sqrt{n!}\sqrt{\frac{m!}{n!}}\hat{\Gamma}_k^{k-1}\sqrt{\frac{(k-1)!}{m!}}|k-1\rangle\\
&=\frac{1}{(k-1)!^{3/2}}\langle 0|\hat{\Gamma}_k^{k-1}|k-1\rangle\\
&=\frac{1}{(k-1)!^{3/2}}\langle 0|(k-1)!^{3/2}|0\rangle\\
&=1.
\end{split}
\end{equation}

Next, we show that all other elements in the upper-left $k\times k$ block $\langle j \vert \hat{P}_{nm} \vert l \rangle=0$, when $j=0,1,\dots,n-1,n+1,\dots,k-1$ and $l=0,1\dots,m-1,m+1,\dots, k-1$. 

We consider three cases: (a) $j<n$; (b) $l<m$, and (c) $l>m$, as follows:

\textit{(a)}. When $j<n$, $\langle j|(a^\dagger)^n=0$. Therefore,
\begin{equation}\label{eq:jl0}
\begin{split}
\langle j|\hat{P}_{nm}|l\rangle&=\langle j|\frac{1}{(k-1)!^2}\sqrt{\frac{m!}{n!}}(\hat{a}^\dagger)^n \hat{\Gamma}_k^{k-1} (\hat{a}^\dagger)^{k-1-m}|l\rangle=0.
\end{split}
\end{equation}

\textit{(b)}. When $l<m$, $\langle j|\hat{P}_{nm}|l\rangle=0$, since $\hat{\Gamma}_k^{k-1}(a^\dagger)^{k-1-m}|l\rangle \propto \hat{\Gamma}_k^{k-1}|l+k-1-m\rangle$, and then according to Eq.~(\ref{eq:gam_a}), $\hat{\Gamma}_k^{k-1}|l+k-1-m\rangle=0$, since $l-m<0$. 

\textit{(c)}. When $l>m$, we obtain $(a^\dagger)^{k-1-m}|l\rangle\propto|k-1-m+l\rangle$. So, according to Eq.~(\ref{eq:gam_a}), $\hat{\Gamma}_k^{k-1}(a^\dagger)^{k-1-m}|l\rangle=0$ since $k-1<k-1-m+l<2k-1$, and $\hat{\Gamma}_k^{k-1} |j\rangle 
=(k-j)(k-(j-1)) \cdots (k-(j-(k-2))) \sqrt{j (j-1) \cdots (j-(k-2))}|j-(k-1)\rangle = 0$ when $j=k,k+1, \dots, 2k-2$ since $(k-j)(k-(j-1)) \cdots (k-(j-(k-2)))=0$. 

To establish block-diagonality, we next show that  $\hat{P}_{nm}$ vanish when: (d) $j\leq k-1$ and $l>k-1$, and also when (e) $l\leq k-1$ and $j>k-1$, as follows: 

\textit{(d). $j\leq k-1$ and $l>k-1$}. This case is further divided into two scenarios: (i) $l<k+m$, or (ii) $l\geq k+m$, as follows: 

(i) $l<k+m$. Similarly to case (c), here $\hat{\Gamma}_k^{k-1}(a^\dagger)^{k-1-m}|l\rangle=0$ since $k-1-m+l<2k-1$. Therefore, $\langle j \vert \hat{P}_{nm}\vert l \rangle=0$.

(ii) $l\geq k+m$. In this case, according to Eq.~\eqref{eq:jl0},
and Eq.~(\ref{eq:Gam_0}), 
\begin{equation}
\begin{split}
\langle j|\hat{P}_{nm}|l\rangle&=\langle j|\frac{1}{(k-1)!^2}\sqrt{\frac{m!}{n!}}(\hat{a}^\dagger)^n \hat{\Gamma}_k^{k-1} (\hat{a}^\dagger)^{k-1-m}|l\rangle\\
&\propto \langle j-n|\hat{\Gamma}_k^{k-1}|l+k-1-m\rangle\\
&\propto \langle j-n|l-m\rangle. 
\end{split}
\end{equation}
Considering that $l\geq k+m$, and $k-1\geq j$, we obtain $l-m\geq k \geq j+1 > j-n$, so
\begin{equation}
\langle j|\hat{P}_{nm}|l\rangle\propto\langle j-n|l-m\rangle=0.
\end{equation}

\textit{(e). $l\leq k-1$ and $j>k-1$}. The argument is analogous to that for case $(d)$. 

Considering cases (a)--(e), we obtain the matrix representation for $\hat{P}_{nm}$ given Eq.~\eqref{eq:Gamma_mn_mat}. Since $n$ and $m$ can be any integer from $0$ to $k-1$, we prove Eq.~\eqref{eq:Hmnmat}. 

\section{Implementing $\hat{R}_z$ and $\hat{R}_x$ with a SNAIL}\label{sec:onequbit}
This section shows that any 1-qubit rotation on the surface of the Bloch sphere can be implemented by using a SNAIL device, introduced in Eq.~\eqref{eq:haman4}, thus enabling a universal set of 1-qubit gates. 

The rotation around the $z$ axis by $\lambda$ has the following matrix representation:
\begin{equation}\label{eq:Rz}
\hat{R}_z(\lambda)=\begin{pmatrix}
1&0\\
0&e^{i\lambda}
\end{pmatrix},
\end{equation}
which can be implemented as $\hat{R}_z(\lambda)=e^{-i \hat{H}_z t}$ by propagating for time $t=1$ a quantum circuit with the effective Hamiltonian,
\begin{equation}\label{eq:Hz}
\hat{H}_z(\lambda)=\begin{pmatrix}
0&0\\
0&-\lambda
\end{pmatrix}.
\end{equation}
Implementing Eq.~\eqref{eq:Hz} with Eq.~\eqref{eq:haman4} requires $\lambda<0$, which correspond to negative rotation angles along the z axis. Noting that any positive rotation angle $\lambda'$ with $0<\lambda'<2\pi$ is equivalent to the negative rotation angle $\lambda=-2\pi+\lambda'$, we show that any rotation around the z axis can be implemented according to Eq.~\eqref{eq:haman4}.

The rotation around the $x$ axis by $\theta$ has the following matrix representation:
\begin{equation}\label{eq:Rx}
\hat{R}_x(\theta)=\begin{pmatrix}
\text{cos}(\theta/2)&-i\text{sin}(\theta/2)\\
-i\text{sin}(\theta/2)&\text{cos}(\theta/2)
\end{pmatrix},
\end{equation}
which correspond to the effective Hamiltonian:
\begin{equation}\label{eq:Hx}
\hat{H}_x(\theta)=\begin{pmatrix}
0&\theta/2\\
\theta/2&0
\end{pmatrix}.
\end{equation}
Mapping Eq.~\eqref{eq:Hx} into Eq.~\eqref{eq:haman4} requires $\omega=0$ --{\em i.e.}, elimination of the linear component by tuning the magnetic flux such that the linear inductance is cancelled out. 

\section{Controlled-Z Gates with Quartic Elements}
\label{sec:quarton}
This section follows and expands Ref.~[\citenum{PhysRevA.59.2631}] to show that the cross-Kerr Hamiltonian,
\begin{equation}
\hat{H}_{cross-Kerr}=\chi\hat{b}_1^\dagger\hat{b}_1\hat{b}_2^\dagger\hat{b}_2,
\end{equation}
implemented with a nearly-quartic element, corresponds to a controlled-Z gate in the basis of Fock states $|0\rangle$ and $|1\rangle$. We show that $e^{-i\pi\hat{b}_1^\dagger\hat{b}_1\hat{b}_2^\dagger\hat{b}_2}$ keeps the basis states $|0\rangle|0\rangle$, $|0\rangle|1\rangle$ and $|1\rangle|0\rangle$ unchanged, while introducing a phase shift of $-1$ to state $|1\rangle|1\rangle$. 

We apply
$e^{-i\hat{H}_{cross-Kerr}t}$ to the outer product states $|0\rangle|0\rangle$, $|0\rangle|1\rangle$, $|1\rangle|0\rangle$, and $|1\rangle|1\rangle$ with $t=\pi/\chi$, so that $e^{-i\hat{H}_{cross-Kerr}t}=e^{-i\pi\hat{a}^\dagger\hat{a}\hat{b}^\dagger\hat{b}}$.

Applying $e^{-i\pi\hat{b}_1^\dagger\hat{b}_1\hat{b}_2^\dagger\hat{b}_2}$ to $|0\rangle|0\rangle$, we obtain:
\begin{equation}
\begin{split}
e^{-i\pi\hat{b}_1^\dagger\hat{b}_1\hat{b}_2^\dagger\hat{b}_2}|0\rangle|0\rangle&=e^{-i\pi \cdot 0\cdot 0}|0\rangle|0\rangle,\\
&=|0\rangle|0\rangle.
\end{split}
\end{equation}

Similarly, 
\begin{equation}
\begin{split}
e^{-i\pi\hat{b}_1^\dagger\hat{b}_1\hat{b}_2^\dagger\hat{b}_2}|0\rangle|1\rangle&=e^{-i\pi \cdot 0\cdot 1}|0\rangle|1\rangle,\\
&=|0\rangle|1\rangle,
\end{split}
\end{equation}

and
\begin{equation}
\begin{split}
e^{-i\pi\hat{b}_1^\dagger\hat{b}_1\hat{b}_2^\dagger\hat{b}_2}|1\rangle|0\rangle&=e^{-i\pi \cdot 1\cdot 0}|1\rangle|0\rangle,\\
&=|1\rangle|0\rangle.
\end{split}
\end{equation}

Finally, 
\begin{equation}
\begin{split}
e^{-i\pi\hat{b}_1^\dagger\hat{b}_1\hat{b}_2^\dagger\hat{b}_2}|1\rangle|1\rangle&=e^{-i\pi \cdot 1\cdot 1}|1\rangle|1\rangle,\\
&=-|1\rangle|1\rangle.
\end{split}
\end{equation}

Therefore, $e^{-i\pi\hat{b}_1^\dagger\hat{b}_1\hat{b}_2^\dagger\hat{b}_2}$ is the controlled-Z gate in the Fock state basis. 

\section{Propagation Method}\label{sec: Popu only}
We compare simulations based on the SBM mapping Hamiltonian, introduced by Eq.~(\ref{Hmn2}), and simulations of quantum dynamics based on the Hamiltonian in the diabatic basis set, introduced by Eq.~(\ref{Hmn}) for the spin-boson model system where $H_{jk}$ is defined, as follows:
\begin{equation}
\begin{split}
\hat{H}_{00} &= \epsilon_{sb} + \sum_{k = 1}^{N_n} \frac{\hat{P}_k^2}{2} + \frac{1}{2}\omega_k^2\hat{R}_k^2 -c_k \hat{R}_k,
\\\hat{H}_{11}  &= -\epsilon_{sb} + \sum_{k = 1}^{N_n} \frac{\hat{P}_k^2}{2} + \frac{1}{2}\omega_k^2\hat{R}_k^2 +c_k \hat{R}_k,
\\ 
H_{01}&= H_{10}  = \Delta_{sb}.
\end{split} 
\label{eq:SBham}
\end{equation}  
The model Hamiltonian, introduced by Eq.~(\ref{eq:SBham}), describes a vibronic system with two electronic states with energy gap $2\epsilon_{sb}$, coupled with each other by the constant coupling constant $\Delta_{sb}$. Each electronic state is coupled to a bath of $N_n$ nuclear degrees of freedom, modeled as displaced harmonic oscillators. For the $k^{th}$ oscillator, the frequency $\{\omega_k\}$ and electron-phonon coupling coefficient, $\{c_k\}$ of the nuclear modes is sampled from an Ohmic spectral density with an exponential cutoff:
\begin{align}
  J (\omega) &= \frac{\pi}{2} \sum_{k=1}^{N_n} \frac{c_k^2}{\omega_k} \delta(\omega-\omega_k) ~  \stackrel{\raisebox{1pt} {\text{\footnotesize$N_n \rightarrow \infty$}}}{\xrightarrow{\hspace*{0.75cm}}} ~ \frac{\pi\hbar}{2}
 \xi \omega e^{-\omega/\omega_c}.
 \label{ohmic}
\end{align}
Here, $\xi$ is the Kondo parameter, which determines the electron-phonon coupling strength, and $\omega_c$ is the cutoff frequency which determines the characteristic vibrational frequency. Therefore, a discrete set of $N_n$ nuclear mode frequencies, $\{\omega_k\}$, and coupling coefficients, $\{c_k\}$, are sampled from the spectral density, introduced by Eq.~(\ref{ohmic})~\cite{Mulvihill2019}. 

\begin{table}
\centering
\caption{{\bf Spin-Boson Model and Simulation Parameters}}
\begin{tabular}{|c|c|c|c|c|c|c|c|}
\hline  $\epsilon_{sb}$ & $\Delta_{sb}$ & $\beta$ & $\xi$ & $\omega_c$  &\ $\omega_{\text{max}}$\ \ & \ \ $N_n$\ \ & $\Delta t$  
  \\ \hline
  1.0   & 1.0  & 5.0  & 0.1  & 1.0  & 5   & 60 & 1.50083 $\times 10^{-3}$
  \\ \hline
\end{tabular} 
\label{tab:parameters}
\end{table}


The initial density matrix $\hat{\rho}(0)$ is assumed to be in the single-product form $\hat{\rho}(0)=\hat{\sigma}(0)\otimes\hat{\rho}_{n}(0)$, where $\hat{\sigma}(0)$ denotes the reduced, electronic density operator written as a $2\times 2$ matrix, and $\hat{\rho}_{n}(0)$, the initial bath density operator, is assumed to be in thermal equilibrium. 

For comparison with benchmark calculations, we obtain the numerically exact time-evolved density matrix $\hat{\rho}(t)$ by propagating the initial density matrix with the numerically exact Tensor-Train Thermo-Field Dynamics (TT-TFD) propagator\cite{Borrelli2016,Borrelli2021,Lyu2023}:
\begin{equation}
\hat{\rho}(t)=e^{-i\hat{H}t}\hat{\rho}(0)e^{i\hat{H}t}.
\end{equation}

Having computed $\hat{\rho}(t)$, we obtain the electronic density operator $\hat{\sigma}(t)=\text{Tr}_n[\hat{\rho}(t)]$ by tracing out the nuclear degrees of freedom. With $\hat{\sigma}(0)$ initialized according to different electronic distributions, and with their corresponding $\hat{\sigma}(t)$ propagated with TT-TFD, we obtain the Liouville space superoperator $\cal{P}$. 
Next we show how to reduce the dimensionality of the non-unitary time evolution super-operator of the spin-boson model to obtain the population-only super-operator as in Eq.~\eqref{vmatvec}.
We note that for the full time evolution operator,
\begin{equation}\label{eq:full_G}
\sigma_{jj}^{\text{full}}(t) = \sum_{l,m = 1}^{N_e} {\cal G}_{jj,lm}^{\text{full}}(t)\sigma_{lm}^{\text{full}}(0).
\end{equation}
When the initial state is diagonal ({\em i.e.}, $\sigma_{jk} (0)= 0$ for $k \neq j$), Eq.~\eqref{eq:full_G} can be simplified, as follows:
\begin{equation}
\sigma_{jj}^{\text{full}}(t) = \sum_{l = 1}^{N_e} {\cal G}_{jj,ll}^{\text{full}}(t)\sigma_{ll}^{\text{full}}(0).
\end{equation}
For the populations-only propagator,
\begin{equation}
\sigma_{jj}^{\text{pop}}(t) = \sum_{l = 1}^{N_e} {\cal G}_{jj,ll}^{\text{pop}}(t)\sigma_{ll}^{\text{pop}}(0).
\end{equation}
Because $\sigma_{jj}^{\text{full}}(t)$ must be equal to $\sigma_{jj}^{\text{pop}}(t)$ \emph{when exact input methods are used}, we can set the right hand sides equal to each other, as follows:
\begin{equation}
\begin{split}
\sum_{l = 1}^{N_e} {\cal G}_{jj,ll}^{\text{full}}(t)\sigma_{ll}^{\text{full}}(0) &= \sum_{l = 1}^{N_e} {\cal G}_{jj,ll}^{\text{pop}}(t)\sigma_{ll}^{\text{pop}}(0)
\\ {\cal G}_{jj,00}^{\text{full}}(t)\sigma_{00}(0) + {\cal G}_{jj,11}^{\text{full}}(t)\sigma_{11}(0) + ... &= {\cal G}_{jj,00}^{\text{pop}}(t)\sigma_{00}(0) + {\cal G}_{jj,11}^{\text{pop}}(t)\sigma_{11}(0) + ...  .
\end{split}
\end{equation}
Therefore, ${\cal G}_{jj,kk}^{\text{full}}(t) = {\cal G}_{jj,kk}^{\text{pop}}(t)$.

For the spin-boson model, ${\cal G}^{\text{pop}}(t)$ is a $2\times 2$ time-dependent matrix.
To obtain the populations-only ${\cal G}^{\text{pop}}(t)$ matrix, we can extract the four corner elements of ${\cal G}^{\text{full}}(t)$:
\begin{equation}
\left(\begin{array}{cccc} 
    {\color{blue}{\cal G}^{\text{full}}_{00,00}(t)} & {\cal G}^{\text{full}}_{00,01}(t) & {\cal G}^{\text{full}}_{00,10}(t) & {\color{magenta}{\cal G}^{\text{full}}_{00,11}(t)}
    \\ {\cal G}^{\text{full}}_{01,00}(t) & {\cal G}^{\text{full}}_{01,01}(t) & {\cal G}^{\text{full}}_{01,10}(t) & {\cal G}^{\text{full}}_{01,11}(t)
    \\ {\cal G}^{\text{full}}_{01,00}(t) & {\cal G}^{\text{full}}_{10,01}(t) & {\cal G}^{\text{full}}_{10,10}(t) & {\cal G}^{\text{full}}_{10,11}(t)
    \\ {\color{orange}{\cal G}^{\text{full}}_{11,00}(t)} & {\cal G}^{\text{full}}_{11,01}(t) & {\cal G}^{\text{full}}_{11,10}(t) & {\color{ForestGreen}{\cal G}^{\text{full}}_{11,11}(t)}
\end{array}\right) 
\Longrightarrow 
\left(\begin{array}{cc} 
    {\color{blue}{\cal G}^{\text{pop}}_{00,00}(t)} & {\color{magenta}{\cal G}^{\text{pop}}_{00,11}(t)}
    \\ {\color{orange}{\cal G}^{\text{pop}}_{11,00}(t)} & {\color{ForestGreen}{\cal G}^{\text{pop}}_{11,11}(t)}
\end{array}\right)
\end{equation}
In this model, the electronic populations can be propagated using the four corner elements of ${\cal G}(t)$, as follows:
\begin{equation}
\left(\begin{array}{c} \sigma_{11}(t) \\ \sigma_{22}(t) \end{array}\right)
= \left(\begin{array}{cc} {\cal G}_{11,11}(t) & {\cal G}_{11,22}(t) \\ {\cal G}_{22,11} & {\cal G}_{22,22}(t) \end{array}\right)
\left(\begin{array}{c} \sigma_{11}(0) \\ \sigma_{22}(0) \end{array}\right).
\end{equation}



\section{Code availability}
The python code for the SBM-SNAIL simulation of the dynamics for the FMO 4-site model is available at: https://github.com/NingyiLyu/SBM-mapping.
\bibliography{References.bib}

\end{document}